\documentclass[11pt,epsf,letterpaper]{article}%
\usepackage[usenames,dvipsnames,svgnames,table]{xcolor}
\usepackage{color}
\usepackage{amsmath}
\usepackage{amsfonts}
\usepackage{verbatim}
\usepackage{amssymb}
\usepackage{graphicx}
\usepackage{epstopdf}
\usepackage{mathrsfs}
\usepackage{xcolor}
\usepackage{cite}%
\providecommand{\U}[1]{\protect\rule{.1in}{.1in}}

\begin{document}

\title{Holographic equation of state in fluid/gravity duality}
\author{$^{(1,2)}$Andr\'{e}s Anabal\'{o}n, $^{(3)}$Dumitru Astefanesei and $^{(4,5)}%
$Robert Mann\\\textit{$^{(1)}$Departamento de Ciencias, Facultad de Artes Liberales,}\\\textit{Universidad Adolfo Ib\'{a}\~{n}ez, Vi\~{n}a del Mar, Chile.}\\\textit{$^{(2)}$DISAT, Politecnico di Torino,}\\\textit{Corso Duca degli Abruzzi 24, I-10129 Torino, Italia.}\\\textit{$^{(3)}$Instituto de F\'\i sica, Pontificia Universidad Cat\'olica de
Valpara\'\i so,} \\\textit{ Casilla 4059, Valpara\'{\i}so, Chile.}\\\textit{$^{(4)}$ Department of Physics and Astronomy, University of Waterloo,}\\\textit{Waterloo, Ontario, Canada N2L 3G1.}\\\textit{$^{(5)}$ Perimeter Institute, 31 Caroline Street North Waterloo,
Ontario Canada N2L 2Y5.}}
\maketitle

\begin{abstract}
We establish a precise relation between mixed boundary conditions for scalar
fields in asymptotically anti de Sitter spacetimes and the equation of state
of the dual fluid. We provide a detailed derivation of the relation in the
case of five bulk-dimensions for scalar fields saturating the
Breitenlohner-Freedman bound. As a concrete example, we discuss the five
dimensional scalar-tensor theories describing a constant speed of sound.

\end{abstract}

It is now widely accepted that there is a precise correspondence
between observables in a $(D-1)$-dimensional gauge field theory and
a $D-$dimensional gravity theory. Indeed, since this duality was
precisely formulated by Maldacena \cite{Maldacena:1997re}, there has
been an increasing amount of conceptual understanding of its
meaning. In particular, the long wavelength regime of the duality,
known as the fluid/gravity correspondence, has attracted much
attention, e.g. \cite{Bhattacharyya:2008jc, Policastro:2001yc,
Janik:2005zt, Bhattacharyya:2007vs}. It follows from the fact that a
relativistic form of the Navier-Stokes equations governing the
hydrodynamic limit of a field theory in $(D-1)$ dimensions, on a
fixed background $\gamma_{ab}$, is equivalent to the dynamics of
$D-$dimensional Einstein gravity with a negative cosmological
constant with $\gamma_{ab}$ as its conformal boundary.

This correspondence allows one to pick a fluid dynamical solution, with an
equation of state dictated by the tracelessness of the boundary energy
momentum tensor, and reconstruct a bulk solution of the full Einstein
equations (for a review see \cite{Hubeny:2011hd}). This opened the possibility
of modeling fluid dynamics using general relativity. For instance the elusive
description of turbulence has been considered within the fluid/gravity
duality, and it has been proposed that gravitational dynamics can become
turbulent when its dual fluid is at large Reynolds number
\cite{Carrasco:2012nf,Adams:2013vsa,Green:2013zba}. This has led to the
definition of a \textquotedblleft gravitational Reynolds
number\textquotedblright\ constructed in terms of the black hole quasinormal
modes \cite{Yang:2014tla}.

The problem with this description is that, in asymptotically anti de
Sitter (AdS) spacetimes with a flat boundary, it is intrinsically
limited to traceless energy momentum tensors and so the actual fluid
is a very particular one. For understanding realistic fluids by
using the fluid/gravity duality, one must be able to describe their
dynamics by a general equation of state that is experimentally
determined. The holographic relation between the real world systems,
which are not conformally invariant in the ultraviolet (UV), and
gravity, requires that the conformal symmetry in the boundary should
be broken.

The main goal of this letter is to propose a concrete relation between the
equation of state of a (non-conformal) fluid and the asymptotic fall-off
behaviour of a scalar field in the AdS bulk. We treat the case of a single
scalar field with mass $m^{2}=-4l^{-2},$ which is the mass of some of the
scalars of type IIB supergravity on $AdS_{5}\times S^{5}$. This case is also
interesting because the mass saturates the Breitenlohner-Freedman
(BF) bound in five dimensions \cite{Breitenlohner:1982bm, BF} and so
the logarithmic branches exist \cite{Bianchi:2001de, Bianchi:2001kw,
Henneaux:2004zi}. We find that, in general, the coefficients of the leading
terms in the asymptotic expansion of a scalar field in AdS gravity determine
the relationship between the pressure and density of a perfect fluid on the
conformal boundary.

This can be traced back to the existence, in any dimension, of two
normalizable modes for scalar fields with masses $m^{2}$ in the window
\cite{Ishibashi:2004wx}
\begin{equation}
-\frac{(D-1)^{2}}{4l^{2}}=m_{BF}^{2}\leq m^{2}<m_{BF}^{2}+l^{-2}\text{ .}
\label{window}%
\end{equation}
From the existence of two normalizable modes it follows that these theories
admit mixed boundary conditions; most of them break the conformal invariance
at the boundary. This implies that the dual energy momentum tensor is not
traceless \cite{Papadimitriou:2007sj, Anabalon:2015xvl} and so the
hydrodynamic limit of the field theory is described by a non-conformal fluid.
We are going to obtain a general holographic equation of state for a time
dependent scalar field using a counterterm method similar in spirit with the
one in \cite{Anabalon:2015xvl} (the work of \cite{Papadimitriou:2007sj} is
based on the Hamilton-Jacobi equation). It follows that our results are easily
generalizable to any dimension and theories with scalars satisfying
(\ref{window}).

Interest in the holographic descriptions of arbitrary equations of
state emerged naturally some years after the AdS/CFT correspondence
was proposed. However, the proposals are restricted to the idea that
one scalar field potential allows the description of one equation of
state \cite{Gursoy:2007cb, Gursoy:2007er, Gubser:2008ny,
Gubser:2008yx,He:2011hw}. The main result of this paper is that
every scalar field potential describes an infinite number of
equations of state. Namely, given a single scalar field potential,
every boundary condition that provides a regular black hole
corresponds to an equation of state. Indeed, when the scalar field
mass is in the BF window (\ref{window}) it provides two integration
constants ($\alpha,\beta$) to the system. A static metric provides
one extra integration constant, $\mu$. So, asymptotically, the
solution space of the fully back-reacted metric plus the scalar
field is characterized by three integration constants
\cite{Liu:2015tqa}. Requiring the existence of a regular horizon
implies that the bulk configuration is completely determined by two
parameters. Integrating the system from the horizon to infinity
constrains the boundary data such that a codimension-1 surface
contains black hole solutions. Schematically, the solution of the
non-linear system of differential equations is the map
\begin{equation}
\alpha=\alpha(\phi_{h},\mathcal{A})\text{,\qquad}\beta=\beta(\phi
_{h},\mathcal{A})\text{,\qquad}\mu=\mu(\phi_{h},\mathcal{A})\text{,}
\label{map}%
\end{equation}
where $\left(  \phi_{h},\mathcal{A}\right)  $ are the horizon data, namely the
value of the scalar field at the horizon, $\phi_{h}$, and the normalized black
hole area\footnote{The normalized area is the total area divided by the
unit-radius area. This will save us some factors in the body of the paper.},
$\mathcal{A}$, and $\left(  \alpha,\beta,\mu\right)  $ are the boundary data.

Indeed, a classical field theory is well defined when the field equations and
boundary conditions are provided, which correspond to a relation of the form
$\beta=\beta(\alpha)$. Therefore, if one is given a theory defined by a
Lagrangian and some boundary condition, knowledge of (\ref{map}) provides an
immediate answer to the question on the existence of black holes for the given
boundary condition: there will be a black hole whenever the boundary condition
intersects the surface defined by (\ref{map}). This intersection is a line. It
can be characterized by a single parameter that corresponds to the unique
integration constant associated with the static black hole. It turns out that
when the boundary condition is fixed the map (\ref{map}) can be written as%
\begin{equation}
\mu=\mu\left(  \alpha\right)  \text{ },\qquad\beta=\beta(\alpha)\text{ }.
\label{reg}%
\end{equation}
We shall see how the regularity condition (\ref{reg}) is the central object
that connects the gravitational description with the equation of state of the
holographic fluid.

Our conventions are defined by the action principle%
\begin{equation}
I\left[  g,\phi\right]  =\int_{M}d^{5}x\sqrt{-g}\left(  \frac{R}{2\kappa
}-\frac{1}{2}\left(  \partial\phi\right)  ^{2}-V(\phi)\right)  +\frac
{1}{\kappa}\int_{\partial M}K\sqrt{-h}+I_{ct}\text{ ,}%
\end{equation}
where $\kappa=8\pi G$ is the reduced Newton constant in five dimensions. The
potential $V(\phi)$ is required to have at least one local maximum, where it
attains a negative value, so that asymptotically AdS solutions exist. The
counterterms $I_{ct}$ are known from well-established results and render the
action principle finite \cite{Balasubramanian:1999re,Mann:1999pc} and well
defined for mixed boundary conditions on the scalar field
\cite{Anabalon:2015xvl}. We are interested in describing a timelike boundary
and so the induced metric on $\partial M$ is $\ h_{\mu\nu}=g_{\mu\nu}-n_{\mu
}n_{\nu}$. The extrinsic curvature of the surface with metric $h_{\mu\nu}$ is
$2K_{\mu\nu}=\mathcal{L}_{n}h_{\mu\nu}=\nabla_{\mu}n_{\nu}+\nabla_{\nu}n_{\mu
}$ where $n_{\mu}$ is the outwards-pointing normal and $K=h^{\mu\nu}K_{\mu\nu
}.$ We use below the Einstein equations as defined by the relation $G_{\mu\nu
}=\kappa T_{\mu\nu}$,where $G_{\mu\nu}$ is the Einstein tensor and $T_{\mu\nu
}=\partial_{\mu}\phi\partial_{\nu}\phi-g_{\mu\nu}\left[  \frac{1}{2}\left(
\partial\phi\right)  ^{2}+V(\phi)\right]  $. Working in units where the speed
of light and Planck's constant are set to $1$, we consider the class of
metrics%
\begin{equation}
ds^{2}=e^{A}\left(  -fdt^{2}+d\Sigma_{k}\right)  +e^{B}\frac{dr^{2}}{f}\text{
,} \label{metclass}%
\end{equation}
where $d\Sigma_{k}$ is a constant Ricci scalar space, $R\left(  \Sigma
_{k}\right)  =6k$, with $k=\pm1$ or $0$, and volume $\sigma_{k}$. The boundary
metric is $h_{\mu\nu}dx^{\mu}dx^{\nu}=e^{A}\left(  -fdt^{2}+d\Sigma
_{k}\right)  $, and the field theory dual metric, which is related by a
conformal transformation to $h_{\mu\nu}$, is%
\begin{equation}
\gamma_{ab}dx^{a}dx^{b}=-dt^{2}+l^{2}d\Sigma_{k}\text{ .} \label{dual}%
\end{equation}
The counterterms $I_{ct}$ are constructed so that the action principle is
well-posed and to obtain a finite action. The quasilocal formalism of Brown
and York \cite{Brown:1992br} provides a concrete way to compute the action and
stress tensor, from which one can directly obtain the mass of the system.

Let us quickly see how one can gain knowledge of the map (\ref{map}). For
static metrics one can introduce the variables $Z=\frac{d\phi}{dA}\equiv
\dot{A}^{-1}~$and $Y=\left(  \dot{A}\dot{f}\right)  ^{-1}f$. When $k=0$ it is
straightforward to see that the Einstein equations are%

\begin{equation}
\dot{Z}=\frac{\left(  3\dot{V}+4\kappa ZV\right)  \left(  2\kappa
Z^{2}Y-6Y-3Z^{2}\right)  }{12\kappa VZY}\text{ ,}\qquad\dot{Y}=\frac{\left(
3\dot{V}+2\kappa ZV\right)  \left(  2\kappa Z^{2}Y-6Y-3Z^{2}\right)  }%
{6VZ^{2}}\text{ .}%
\end{equation}
It follows that $\dot{Z}$ is finite at the horizon, located at $Y(\phi_{h}%
)=0$, if and only if $Z\left(  \phi_{h}\right)  =-\frac{3\dot{V}\left(
\phi_{h}\right)  }{4\kappa V\left(  \phi_{h}\right)  }$. \ Furthermore, we
readily see that these equations decouple and reduce to the single master
equation%
\begin{equation}
-3\left(  3\dot{V}+4\kappa ZV\right)  Z\ddot{Z}+\left(  -9\dot{V}+12\kappa
ZV\right)  \dot{Z}^{2}+\left[  8\kappa^{2}Z^{3}V+\kappa\left(  24ZV+18Z^{2}%
\dot{V}\right)  +18\dot{V}+9Z\ddot{V}\right]  \dot{Z}=0\text{ .} \label{MEQ}%
\end{equation}
The regularity condition imply that $Z$ is completely determined by the value
of $\phi$ at the horizon, and so
\begin{equation}
A\left(  r\right)  =\int_{\phi_{h}}^{\phi(r)}\frac{d\phi}{Z}+\frac{2}{3}%
\ln\left(  \mathcal{A}\right)  \text{ ,}%
\end{equation}
which follows from the definition of $Z$ and the value of $A(r_{h})$. When the
scalar field saturates the BF bound, $m^{2}=-\frac{4}{l^{2}}$ its fall-off is%
\begin{equation}
\phi=\frac{\alpha\ln(r)}{r^{2}}+\frac{\beta}{r^{2}}+O(\frac{\ln(r)}{r^{3}%
})\text{ }. \label{phi1}%
\end{equation}
It is a consequence of the fall-off of the scalar (\ref{phi1}) that,
asymptotically, $e^{A}\phi\sim\alpha\ln(r).$ The derivative of this relation
with respect to $\ln r\sim\frac{A}{2}$ yields $2e^{A}\left(  \phi+\frac{d\phi
}{dA}\right)  \sim\alpha$. Thus,
\begin{equation}
\alpha\left(  \phi_{h},\mathcal{A}\right)
= {\lim_{\phi\rightarrow 0}}2\left(  Z+\phi\right)
e^{A}=\alpha_{h}\left(  \phi_{h}\right)  \mathcal{A}^{\frac{2}{3}}\text{ .}
\label{map 1}%
\end{equation}
We see that $\alpha$ is generically a function of $\phi_{h}$ times a
very precise function of the normalized black hole area. What is remarkable
here is that, since the map (\ref{map}) is coming from
the complete integration of Einstein equations, one might have expected that
the boundary data is a very complicated function of the horizon data,
but that is not the case. It follows from the straightforward derivation of
(\ref{map 1}) that the same sort of relation exists for scalar fields with
masses in the BF window (\ref{window}).

To construct the dual energy momentum tensor we need to identify the total
energy of the system. We shall use then the Regge-Teilteiboim approach
\cite{Henneaux:2006hk, Regge:1974zd}. We emphasize below that the coefficients
$\alpha$, $\beta$ and $\mu$ can be generalized from integration constant to be
time-dependent. The calculation can be done in full generality with
all the boundary coordinate dependence however, for the sake of simplicity, we
restrict it to time dependence. When the metric matches (locally) $AdS$ at
infinity the relevant fall-off is%
\begin{align}
g_{tt}  &  =\frac{r^{2}}{l^{2}}+k-\frac{\mu(t)}{r^{2}}+O(r^{-3}%
)\,\,\,\,\,\,\,\,\,\,\,\text{ ,}\,\,\,\,\,\,\,\,\,\,\,\,\,\,\,g_{ij}%
=r^{2}\Sigma_{ij}+O(r^{-3})\text{ ,}\label{BC1}\\
g_{rr}  &  =\frac{l^{2}}{r^{2}}-\frac{l^{4}k}{r^{4}}+\frac{l^{2}}{3}%
\frac{3M_{h}(t)l^{2}+3k^{2}l^{4}+\kappa\alpha(t)\left(  \alpha(t)-4\beta
(t)\right)  \ln(r)-2\kappa\alpha(t)^{2}\ln(r)^{2}}{r^{6}}+O\left(  \frac
{\ln(r)^{2}}{r^{7}}\right)  \text{ ,} \label{BC2}%
\end{align}
where $\Sigma_{ij}$ is the metric associated with the \textquotedblleft
angular\textquotedblright\ part, $d\Sigma_{k}$. Inserting these expansions in
the Einstein-scalar field equations we find that the boundary conditions
(\ref{BC1})-(\ref{BC2}) are compatible with the field equations provided
\begin{equation}
M_{h}(t)=\mu(t)-\frac{\kappa l^{-2}}{12}\left(  \alpha(t)^{2}-4\alpha
(t)\beta(t)+8\beta(t)^{2}\right)  \text{ .} \label{grr}%
\end{equation}
Using the fall-off of the metric and scalar field we obtain
\begin{equation}
\delta H=\left[  \frac{3\delta M_{h}(t)}{2\kappa}-\frac{1}{l^{2}}\left(
\alpha(t)\delta\beta(t)-2\beta(t)\delta\beta(t)\right)  \right]  \sigma
_{k}\text{ ,} \label{deltH}%
\end{equation}
and so the Hamiltonian is finite\footnote{The gravitational and scalar
contributions to the Hamiltonian are universal when given in terms of its
variations
\begin{equation}
\delta H=\delta Q_{G}+\delta Q_{\phi}\text{ ,} \label{deltB}%
\end{equation}
and the concrete expressions can be found in \cite{Henneaux:2006hk} --- exact solutions
were studied in \cite{Anabalon:2013qua, Anabalon:2013sra, Anabalon:2013eaa}.
\par
{}}. To remove the variations from these equations we need to impose boundary
conditions on the scalar field. If we write $\beta=\frac{dW(\alpha)}{d\alpha}$
then the right-hand side of (\ref{deltH}) is a total variation and using the
field equations (\ref{grr}) it yields
\begin{equation}
H=\left[  \frac{3\mu(t)}{2\kappa}+\frac{1}{l^{2}}\left(  -\frac{1}{8}%
\alpha(t)^{2}-\frac{1}{2}\alpha(t)\beta(t)+W(\alpha)\right)  \right]
\sigma_{k}+H_{h}\text{ ,} \label{E}%
\end{equation}
where $\mu(t)$ is the $O(r^{-2})$ coefficient of the $g_{tt}$ and $\delta
H_{h}=0$.

It was originally pointed out in \cite{Ishibashi:2004wx} that the evolution of
scalar fields in AdS is well defined for Robin boundary conditions for scalar
fields with masses that satisfy $m_{BF}^{2}\leq m^{2}<m_{BF}^{2}+l^{-2}$ where
$m_{BF}^{2}$ is the Breitenlohner-Freedman bound, $m_{BF}^{2}=-\frac{4}{l^{2}%
}$ \cite{BF}. Indeed, it is possible to find this kind of formula in a number
of places in the literature \cite{Hertog:2004dr}. What is new here is that we
have taken one order more in the fall off of $g_{tt}$, namely the $\mu/r^{2}$
term, and shown how it connects with the standard definition of mass given in
terms of the coefficient of $O(r^{-2})$ of $g_{rr}^{-1}$ (see, also,
\cite{Anabalon:2014fla}).

Due to integration over an infinite volume, the action suffers from infrared
divergences that can be regulated by adding suitable boundary terms. With this
in mind, the action can be naturally divided into the bulk part, the usual
Gibbons-Hawking boundary term, the Balasubramanian-Kraus counterterm, an
extrinsic scalar field counterterm, $I_{ext}^{\phi}$, and an intrinsic scalar
field counterterm, $I_{ct}^{\phi}$:
\begin{equation}
I=I_{B}+I_{GH}+I_{BK}+I_{ext}^{\phi}+I_{ct}^{\phi}\text{ .} \label{TA}%
\end{equation}
The boundary conditions (\ref{phi1}), (\ref{BC1}), (\ref{BC2}) and the field
equations imply that it is possible to introduce the following counterterms
that provide the correct result for the free energy
\[
I_{ext}^{\phi}=\frac{1}{2}\int_{\partial M}d^{4}x\sqrt{-h}n^{\mu}\phi
\partial_{\mu}\phi\text{ ,}\qquad I_{ct}^{\phi}=\frac{1}{l^{5}}\int_{\partial
M_{\gamma}}d^{4}x\sqrt{-\gamma}\left[  \frac{\alpha\beta}{2}-W(\alpha)\right]
\text{ ,}%
\]
where we have used the metric $\gamma_{ab}$ of the dual field theory
description. When the field equations hold, the variation of the total action
(\ref{TA}) vanishes for Dirichlet boundary conditions for the metric and for
scalar field boundary conditions of the form $\beta\left(  t\right)
=\frac{dW}{d\alpha}$, namely
\begin{equation}
\lim_{r\rightarrow\infty}\delta I=0\text{ .} \label{VP}%
\end{equation}
Let us clarify this further for the scalar field. From (\ref{TA}) we obtain
\begin{align}
\delta I  &  =\int_{M}-d^{5}x\partial_{\mu}\left(  \sqrt{-g}g^{\mu\nu}%
\delta\phi\partial_{\nu}\phi\right)  +\frac{1}{2}\int_{\partial M}d^{4}%
x\sqrt{-h}n^{\mu}\delta\phi\partial_{\mu}\phi+\frac{1}{2}\int_{\partial
M}d^{4}x\sqrt{-h}n^{\mu}\phi\partial_{\mu}\delta\phi\nonumber\\
&  +\frac{1}{l^{5}}\int_{\partial M_{\gamma}}d^{4}x\sqrt{-\gamma}\left(
-\frac{\beta\left(  t\right)  }{2}+\frac{\alpha\left(  t\right)  }{2}%
\frac{d^{2}W}{d\alpha^{2}}\right)  \delta\alpha\left(  t\right)  \text{ ,}%
\end{align}
when the field equations hold. Using
\begin{equation}
\phi=\frac{\alpha\left(  t\right)  \ln(r)}{r^{2}}+\frac{\beta\left(  t\right)
}{r^{2}}+O(\frac{\ln(r)}{r^{3}})\Longrightarrow\delta\phi=\left(  \ln
(r)+\frac{d^{2}W}{d\alpha^{2}}\right)  \frac{\delta\alpha\left(  t\right)
}{r^{2}}+O(\frac{\ln(r)}{r^{3}})\text{ ,}%
\end{equation}
and employing (\ref{phi1}), (\ref{BC1}), (\ref{BC2}) and (\ref{grr}), it is
straightforward to show that (\ref{VP}) indeed holds.

There is one remaining ambiguity in (\ref{TA}), namely that of adding finite
counterterms quadratic in the Riemann tensor, Ricci tensor and Ricci scalar of
the boundary metric. This is related to the regularization of the field theory
dual as discussed in \cite{Balasubramanian:1999re}.

\section*{Holographic Smarr formula and equation of state}

The expectation value of the dual energy-momentum tensor is related to the
quasilocal stress tensor (incuding the counterterms):
\begin{equation}
\left\langle \mathcal{T}_{ab}\right\rangle =-\frac{2}{\sqrt{-\gamma}}%
\frac{\delta I}{\delta\gamma^{ab}}=\lim_{r\rightarrow\infty}\frac{r^{2}}%
{l^{2}}\mathcal{T}_{\mu\nu}^{BK}+\lim_{r\rightarrow\infty}\frac{r^{2}}{l^{2}%
}\mathcal{T}_{\mu\nu}^{ext}+\mathcal{T}_{ab}^{ct}\text{ ,} \label{Tensor}%
\end{equation}
where the first term is the Balasubramanian-Kraus part
\cite{Balasubramanian:1999re}%
\begin{equation}
\mathcal{T}_{\mu\nu}^{BK}=-\frac{1}{\kappa}\left(  K_{\mu\nu}-h_{\mu\nu
}K+\frac{3}{l}h_{\mu\nu}-\frac{l}{2}\mathcal{G}_{\mu\nu}\right)  \text{ ,}%
\end{equation}
with $\mathcal{G}_{\mu\nu}$ the Einstein tensor constructed with the metric
$h$. The second term and the third term are the extrinsic scalar field term
and the finite contribution contributions introduced in this paper:%
\begin{equation}
\mathcal{T}_{\mu\nu}^{ext}=\frac{1}{2}h_{\mu\nu}n^{\mu}\phi\partial_{\mu}%
\phi\text{ ,}\qquad\mathcal{T}_{ab}^{ct}=\frac{1}{l^{5}}\gamma_{ab}\left[
\frac{\alpha\left(  t\right)  \beta\left(  t\right)  }{2}-W(\alpha)\right]
\text{ .}%
\end{equation}
The relevant divergence coming from the bulk and Gibbons-Hawking contributions
is canceled out by the divergence from the counterterm and we obtain the
following regularized stress tensor of the dual field theory:
\begin{align}
\left\langle \mathcal{T}_{ab}\right\rangle  &  =\frac{\gamma_{ab}}{l^{3}%
}\left[  -\frac{3M_{h}\left(  t\right)  }{2\kappa}+\frac{k^{2}l^{2}}{8\kappa
}+\frac{2\mu\left(  t\right)  }{\kappa}+\frac{1}{l^{2}}\left(  \alpha\left(
t\right)  \beta\left(  t\right)  -\beta\left(  t\right)  ^{2}-W(\alpha
)\right)  \right] \nonumber\\
&  +\frac{1}{\kappa l}\delta_{a}^{0}\delta_{b}^{0}\left[  \frac{k^{2}}%
{2}+\frac{2\mu\left(  t\right)  }{l^{2}}\right]  \text{ .} \label{BT}%
\end{align}
Taking the trace and using the field equations (\ref{grr}), we obtain
\begin{equation}
\gamma^{ab}\left\langle \mathcal{T}_{ab}\right\rangle =\frac{1}{l^{5}}\left[
\frac{1}{2}\alpha\left(  t\right)  ^{2}+2\alpha\left(  t\right)  \beta\left(
t\right)  -4W(\alpha)\right]  \text{ ,}%
\end{equation}
which vanishes for the AdS invariant boundary conditions. Using the normalized
timelike vector $u^{a}=\partial_{t}$, the energy density of the fluid is
\[
\rho=u^{a}u^{b}\left\langle \mathcal{T}_{ab}\right\rangle =\frac{1}{l^{3}%
}\left[  \frac{3M_{h}\left(  t\right)  }{2\kappa}+\frac{3k^{2}l^{2}}{8\kappa
}-\frac{1}{l^{2}}\left(  \alpha\left(  t\right)  \beta\left(  t\right)
-\beta\left(  t\right)  ^{2}-W(\alpha)\right)  \right]  \text{ }.
\]
The total mass is the energy density integrated on a spacelike section
\begin{equation}
M=\int_{\Sigma}\rho l^{3}d\Sigma=\left[  \frac{3M_{h}\left(  t\right)
}{2\kappa}+\frac{3k^{2}l^{2}}{8\kappa}+\frac{1}{l^{2}}\left(  -\alpha\left(
t\right)  \beta\left(  t\right)  +W(\alpha)+\beta\left(  t\right)
^{2}\right)  \right]  \sigma_{k}=H\text{ ,}%
\end{equation}
where the last equality is to remark that this result is in agreement with the
Hamiltonian with $H_{h}=\frac{3k^{2}l^{2}}{8\kappa}$. The counterterm
computation also provides the Casimir energy of the large N limit of
$\mathcal{N}=4$ Super Yang-Mills theory --- a cross check of our computation
is its exact agreement with the original paper of Balasubramanian-Kraus when
the scalar field vanishes \cite{Balasubramanian:1999re}.

The introduction of the scalar field yields a dual perfect fluid with energy
momentum tensor $\left\langle \mathcal{T}_{ab}\right\rangle =\left(
\rho+p\right)  u_{a}u_{b}+p\gamma_{ab}$. Hence, we can identify%
\begin{equation}
p=\frac{1}{l^{3}}\left[  \frac{\mu\left(  t\right)  }{2\kappa}+\frac
{k^{2}l^{2}}{8\kappa}+\frac{1}{8l^{2}}\left(  \alpha\left(  t\right)
^{2}+4\alpha\left(  t\right)  \beta\left(  t\right)  -8W(\alpha)\right)
\right]  \label{press}%
\end{equation}%
\begin{equation}
\rho=\frac{1}{l^{3}}\left[  \frac{3\mu\left(  t\right)  }{2\kappa}%
+\frac{3k^{2}l^{2}}{8\kappa}-\frac{1}{8l^{2}}\left(  \alpha\left(  t\right)
^{2}+4\alpha\left(  t\right)  \beta\left(  t\right)  -8W(\alpha)\right)
\right]  \label{dens}%
\end{equation}
where we have used the relation (\ref{grr}). Note that when there is
no scalar field we get a thermal gas of massless particles $\rho=3p$
\cite{Myers:1999psa}. When $k=0$, an interesting implication of
(\ref{press}) and (\ref{dens}) is that the entropy density,~$s=$
$\frac{\mathcal{A}}{4G},$ and temperature $T$ of a perfect fluid is
defined by the relation\footnote{The
$l^{3}$ factor is due to our definition of the dual metric (\ref{dual}).}%
\begin{equation}
Ts=l^{3}\left(  \rho+p\right)  =\frac{2\mu\left(  t\right)  }{\kappa}\text{ ,}
\label{Smarr}%
\end{equation}
which exactly coincides with the generalized Smarr formula of
\cite{Liu:2015tqa}. Our calculation shows that the same formula holds when the
gravitational configuration is time dependent. When $k=0,$ the temperature of
the configuration has the form $T=\frac{8G}{\kappa l^{2}}\mu_{h}\left(
\phi_{h}\right)  \mathcal{A}^{\frac{1}{3}}$. Using (\ref{Smarr}) we find
\begin{equation}
\mu\left(  \phi_{h},\mathcal{A}\right)  =\frac{\mu_{h}\left(  \phi_{h}\right)
\mathcal{A}^{\frac{4}{3}}}{l^{2}}\text{ .} \label{mu}%
\end{equation}
Inserting (\ref{map}) in the first law of black hole thermodynamics, $\delta
H=T\delta S$, with the knowledge of (\ref{mu}) and (\ref{map 1}) shows that
the terms proportional $\delta\mathcal{A}$ cancel, provided that
\begin{equation}
\beta\left(  \phi_{h},\mathcal{A}\right)  =-\frac{1}{2}\alpha\ln\left(
\alpha\right)  -\frac{1}{2}\alpha\beta_{h}\left(  \phi_{h}\right)  ~.
\label{beta}%
\end{equation}
Moreover, the terms proportional to the variation of $\delta\phi_{h}$ cancel
if and only if%
\begin{equation}
6d\mu_{h}+\kappa\alpha_{h}^{2}d\beta_{h}=0\text{ .} \label{mu0}%
\end{equation}
By the implicit function theorem we can take the independent variable to be
$\alpha_{h}$ and hence reduce the problem to that of finding only one
function, for instance $\beta_{h}$ obtaining then $\mu_{h}$ by a direct
integration of (\ref{mu0}). Finally, note that fixing the boundary condition
implies a functional relation between $\phi_{h}$ and $\mathcal{A}$. This is
equivalent to say that the black hole is characterized by a single integration constant.

A useful consequence of the general considerations made so far, it that there
are two $\mathcal{A}-$independent functions
\begin{equation}
\frac{\beta}{\alpha}+\frac{1}{2}\ln\left(  \alpha\right)  \text{ },\qquad
\frac{\mu}{\alpha^{2}}\text{ }.\label{Aind}%
\end{equation}
Whenever there is a hairy black hole with AdS invariant conditions at
$\phi_{h}=\phi_{\ast}$ then the functions $\left(  \beta_{h},\alpha_{h}%
,\mu_{h}\right)  $ admit a Taylor expansion around $\phi_{h}=\phi_{\ast
}\footnote{The numerical integration of the field equations done in
\cite{Hertog:2004dr} shows that for a scalar field belonging to a consistent
truncation of a type IIB supergravity Lagrangian it is possible to obtain black
holes with AdS invariant boundary conditions for any finite constant value of
$\beta_{h}$.}$. It follows then that one can obtain the generic form of the
surface of existence of hairy black holes around a given regular point
\begin{equation}
\frac{\beta}{\alpha}+\frac{1}{2}\ln\left(  \alpha\right)  =-\frac{1}{2}%
\beta_{\ast}+\frac{C}{\kappa}\left(  \frac{\mu l^{2}}{\alpha^{2}}-\frac
{\mu_{\ast}}{\alpha_{\ast}^{2}}\right)  +O\left(  \frac{\mu^{2}}{\alpha^{4}%
}\right)  \text{ ,}\label{eq1}%
\end{equation}
where it was used that exists a hairy black hole with AdS invariant boundary
conditions at $\left(  \beta_h,\alpha_h,\mu_h\right)  =\left(  \beta
_\ast,\alpha_\ast,\mu_\ast\right)  $ and $C$ is a constant that depends on the
theory. When higher order corrections are neglected, insertion of
equation (\ref{eq1}) in (\ref{mu0}) yields
\begin{equation}
\mu_{h}=\mu_{\ast}\left(  \frac{\alpha_{h}}{\alpha_{\ast}}\right)  ^{\frac
{2C}{-3+C}},\qquad\beta_{h}=\beta_{\ast}-\frac{2C}{\kappa}\frac{\mu_{\ast}%
}{\alpha_{\ast}^{2}}\left(  \left(  \frac{\alpha_{h}}{\alpha_{\ast}}\right)
^{\frac{6}{-3+C}}-1\right)  \text{ .}\label{eq1a}%
\end{equation}
upon using \eqref{beta}, the integration constant being fixed by
requiring $\alpha_{h}=\alpha_{\ast
}\Longrightarrow\mu_{h}=\mu_{\ast}$.

We have now the tools to further analyze the connection between infrared
regularity and the equation of state. When the boundary conditions are fixed,
$\beta=\frac{dW(\alpha)}{d\alpha}$, the density and pressure are specified by
the choice of $W(\alpha)$. This is tantamount to defining an equation of
state. Conversely, specification of an equation of state $p=p(\rho)$
necessarily determines $W(\alpha)$ from equations (\ref{press}) and
(\ref{dens}). Indeed, we need the function $\mu(\alpha)$ to determine the
exact boundary condition associated to a given equation of state of the dual
fluid. However, we will keep the discussion general and treat a simple case.

For instance, we find that the equation of state $p=c_{s}^{2}\rho$ where
$c_{s}^{2}$ is the (constant) speed of sound squared, is equivalent to the
following one-parameter family of boundary conditions
\begin{equation}
W(\alpha)=\frac{l^{2}\left(  3c_{s}^{2}-1\right)  }{\kappa\left(  c_{s}%
^{2}+1\right)  }\left(  \alpha^{2}\omega\left(  \alpha\right)  -\frac
{k^{2}l^{2}}{8}\right)  -\frac{\alpha^{2}}{4}\left(  \ln(\alpha)+\alpha
_{1}-\frac{1}{2}\right)  \text{ ,} \label{1param}%
\end{equation}
where we have parameterized $\mu=\alpha^{3}\frac{d\omega\left(  \alpha\right)
}{d\alpha}$ and $\alpha_{1}$ is an integration constant. We readily see that
when $c_{s}^{2}=\frac{1}{3}$ we recover the description of the gas of massless
particles and the AdS invariant boundary conditions. Now, let us assume that
there is a black hole with AdS invariant boundary conditions at $\alpha
_{1}=\beta_{\ast}$. Then, we can use (\ref{eq1}) to find $\mu(\alpha)$ and
finally to find $\beta(\alpha):$%

\begin{equation}
\beta(\alpha)=\alpha\left[  \left(  -\frac{1}{2}+\frac{3\mu_{\ast}\left(
3c_{s}^{2}-1\right)  }{2\kappa\alpha_{\ast}^{2}}\right)  \ln(\alpha
)-\frac{\beta_{\ast}}{2}-\frac{\mu_{\ast}\left(  3c_{s}^{2}-1\right)  \left(
C-3\right)  }{4\kappa\alpha_{\ast}^{2}}\right]  +O\left(  \left(  3c_{s}%
^{2}-1\right)  ^{2}\right)  \label{BC}%
\end{equation}
where we linearize around $c_{s}^{2}=\frac{1}{3}$ to be consistent with the
fact that we have neglected $O\left(  \frac{\mu^{2}}{\alpha^{4}}\right)  ~$in
(\ref{eq1}). The dependence of the boundary condition (\ref{BC}) on $C$ shows
that for any theory defined by the surface (\ref{eq1}) one can find the
corresponding boundary condition that yields the desired equation of state.
This is valid in a neighborhood of the AdS invariant boundary condition.
Hence, it is valid for equations of state that can be made arbitrarily close
to the gas of massless particles.

Although we have worked in five dimensions for scalar fields saturating the
Breitenlohner-Freedman bound, our results are easily generalized to any
spacetime dimension for any scalar field with masses between this bound and
the unitarity bound. This can be of particular use to the holographic
description of metals, superconductors and different kind of materials
\cite{Hartnoll:2008vx}. The holographic description of condensed matter
systems has recently been discussed in the hydrodynamic regime
\cite{Davison:2016hno}. The results bringed in here allow to actually
introduce a detailed description of the condensed matter system through its
equation of state, in the holographic picture.

The formalism introduced here has a direct application on the exact, time
dependent hairy black hole solutions in Einstein-dilaton gravity with general
moduli potential, recently constructed in \cite{Zhang:2014sta, Lu:2014eta,
Zhang:2014dfa, Fan:2015tua, Fan:2015ykb}. Indeed, all these collapsing black
holes are dual to some process in fluid/gravity with a very precise equation
of state that can now be unveiled.

We have seen that contrary to the common belief, there are many dual fluid
equations of state associated to one theory. This is particularly relevant for
string theory. The construction of the map (\ref{map}) along the lines
described in this letter for type IIB supergravity will provide an holographic
description of the fluid dynamics associated to the deformations of
$\mathcal{N}=4$ super Yang-Mills well behaved in the infrared.

\section*{Acknowledgments}

Research of AA is supported in part by Fondecyt Grant 1141073 and
Newton-Picarte Grants DPI20140053 and DPI20140115. The work of DA is supported
by the Fondecyt Grant 1161418 and Newton-Picarte Grant DPI20140115. This work
was supported in part by the Natural Sciences and Engineering Research Council
of Canada.


\end{document}